\documentclass[letter]{aa}
\usepackage{graphicx}
\usepackage{txfonts}
\usepackage{etoolbox}
\usepackage[hidelinks]{hyperref}
\makeatletter
\patchcmd{\enddocument}{\@enddocumenthook}{\UseOneTimeHook{enddocument}}{}{}
\patchcmd{\aa@pageof}{\pageref{LastPage}}{\pageref*{LastPage}}{}{}
\makeatother

\newcommand{\euclid}{\textit{Euclid}}
\newcommand{\scor}{S_{\rm col}}
\newcommand{\pstar}{P_{\star}}
\begin{document}
\title{Searching for lensed quasars in \euclid\ DR1-Foundation}
\subtitle{I. 26 new candidates from a catalogue-based search}
% Confirm author list, affiliations and acknowledgements before submission.
\author{Zizhao He\inst{1,2}
        \and Qinwen Tang\inst{1,2}
        \and Fuwen Shu\inst{1,2}}
\institute{Department of Physics, Nanchang University, Nanchang, 330031, China\\
           \email{zzhe@ncu.edu.cn}
           \and
           Center for Relativistic Astrophysics and High Energy Physics,
           Nanchang University, Nanchang, 330031, China}
\date{}
\titlerunning{Searching for lensed quasars in \euclid\ DR1-Foundation. I}
\authorrunning{Z. He et al.}
\abstract
{Time delays between the images of a strongly lensed quasar constrain the Hubble constant when combined with a lens mass model. In four-image systems, flux ratios probe dark matter substructure in the lens galaxy. The full \euclid\ Wide Survey is expected to identify up to 2300 lensed quasars.}
{Only two of the 23 known lensed quasars in the Wide Survey sample of \euclid\ Data Release~1--Foundation show conspicuous arcs in visible light. We therefore search for new candidates using multiple compact quasar images and assess which known lenses our catalogue selection recovers.}
{From a parent catalogue of 57\,424\,066 photometric quasar candidates, we select pairs separated by $0.1$--$4$~arcsec using four-band colours, stellar probabilities, and a minimum signal-to-noise ratio of 20 per component in each band. We link pairs that share a source into 698 groups and visually inspect the 693 with usable four-band images. We fit the four-band images of the candidates and use their point-source positions to constrain a singular isothermal ellipsoid (SIE) for each system.}
{Of these 23 known lensed quasars, four have at least two quasar images matched to separate entries in the parent catalogue. The search recovers all four reference lenses (4/23 overall, 17.4\%) and yields 26 additional candidates. The median photometric source redshift is $z_{\rm s}=1.81$ across the 26 candidates, using the mean of the available image redshifts for each system. The 12 candidate deflectors with photometric redshifts span $z_{\rm d}=0.33$--$1.12$, with a median of 0.67. The SIE Einstein radii span $\theta_{\rm E}=0.21\arcsec$--$2.64\arcsec$, with a median of $1.01\arcsec$.}
{}
\keywords{gravitational lensing: strong -- quasars: general -- surveys -- catalogs}
\maketitle

\section{Introduction}

Strongly lensed quasars probe cosmic expansion, dark matter, and quasar structure. Assuming a flat universe with a cosmological constant and cold dark matter ($\Lambda$CDM), the Time-Delay Cosmography collaboration combined time-delay distances from lens models \citep{Refsdal1964,Wong2020,Xie2025,Paic2026} with stellar kinematics and a supernova prior on matter density to measure the Hubble constant to 3.3\% precision from 11 lenses \citep{TDCOSMO2026Measurement}. In four-image systems, lens substructure and line-of-sight haloes perturb flux ratios \citep{DalalKochanek2002,Gilman2020}, allowing \textit{James Webb Space Telescope} measurements to constrain dark matter self-interactions and free streaming \citep{Gilman2026SIDM,Gilman2026Arcs}. Stellar microlensing probes accretion-disc size through chromatic continuum magnification \citep[see also \citealt{Zheng2022} for magnification maps]{Anguita2008} and broad-line region size and kinematics through differential magnification across line profiles \citep{Fian2021}.

The \euclid\ Wide Survey could reveal up to 2300 lensed quasars, about 16\% of them quads \citep{Euclid2024}. Lens searches in \euclid\ Data Release~1--Foundation (DR1-Foundation) have focused on extended arcs. Across roughly 2000~deg$^2$ \citep{Aussel2026DR1}, the Euclid Lens Identification Pipeline and Scientific Exploration (ECLIPSE) identified 15\,072 grade~A or B galaxy--galaxy lenses \citep{Lines2026ECLIPSE}. About 300 have active galactic nucleus counterparts. In about 65, either the lensed source or the deflector could host the activity (Sect.~6.3 of \citealt{Lines2026ECLIPSE}). Our search starts from multiple compact quasar images.

Searches for multiple compact images have identified lensed quasars in other wide-area surveys. In Gaia, multiple detections around a quasar candidate provided targets for spectroscopic confirmation \citep{Lemon2023}. In the Dark Energy Spectroscopic Instrument (DESI) Legacy Imaging Surveys, searches for groups of neighbouring quasar candidates yielded a combined sample of 1115 distinct systems \citep{Dawes2023,He2023}. At larger separations, \citet{Wu2026} searched the CatNorth catalogue \citep{Fu2024CatNorth} for groups of quasars with similar colours and maximum separations of $10$--$72\arcsec$.

\begin{figure*}[!t]
\centering
\includegraphics[width=\textwidth]{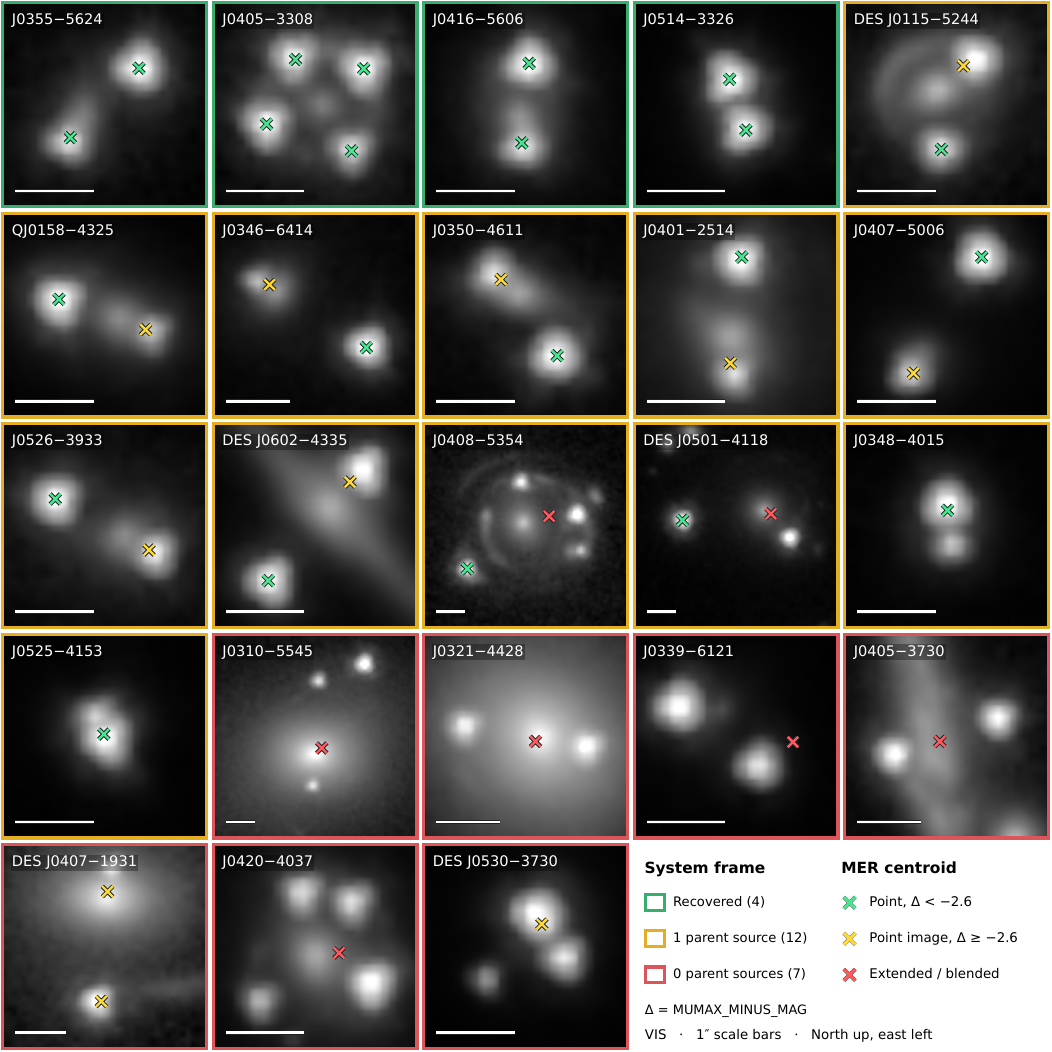}
\caption{VIS images of the 23 GraL lenses with usable Wide Survey imaging from archive release \texttt{DR1\_R2}. Green frames mark the four recovered lenses. Yellow and red frames mark systems with one and zero parent-catalogue sources, respectively.}
\label{fig:recovered}
\end{figure*}

We inspected \euclid\ Visible Instrument (VIS) images of 23 known lensed quasars from the Gaia Gravitational Lenses (GraL) catalogue. The DR1-Foundation Wide images, shown with merged-source (MER) catalogue positions in Fig.~\ref{fig:recovered}, reveal prominent arcs in only two systems (\mbox{J0115$-$5244} and \mbox{J0408$-$5354}). We therefore search for the compact quasar images themselves, applying the catalogue-based method of \citet{He2023} to identify close groups of point sources with similar colours. In Sect.~\ref{sec:data} we describe the four-band data and define the reference lenses used to assess recovery. We describe the selection and the light and mass models in Sect.~\ref{sec:methods}. We report the recovery of all four reference lenses and present 26 additional candidates in Sect.~\ref{sec:results}. In Sect.~\ref{sec:discussion} we examine which known lenses escape selection, how to recover them, and what the results imply for the full Wide Survey. Appendix~\ref{app:catalogue} presents the candidate images and their singular isothermal ellipsoid (SIE) fits. Unless otherwise stated, all magnitudes are on the AB system \citep{OkeGunn1983}.

\section{Data}\label{sec:data}
We select candidate quasi-stellar objects (QSOs) from the \euclid\ catalogue by requiring point-source morphology and low stellar probability. We compare their colours using VIS, $Y$, $J$, and $H$ photometry. Resolved images in the same bands show how the candidate quasar images are arranged around a possible deflector (Sect.~\ref{sec:euclid-data}). Known lensed quasars with multiple counterparts in the parent catalogue provide the reference lenses for the search (Sect.~\ref{sec:reference-lenses}).

\subsection{Euclid data and the Wide QSO parent catalogue}\label{sec:euclid-data}
The \euclid\ Consortium and the European Space Agency (ESA) released DR1-Foundation on 12 November 2026. Its Wide Survey products include images, source catalogues, and spectra covering approximately 1900~deg$^2$ \citep{Aussel2026DR1}.

The Wide Survey imaging covers four optical and near-infrared bands. VIS samples the optical image at $0.1\arcsec$ per detector pixel \citep{Cropper2025VIS}, and the Near-Infrared Spectrometer and Photometer samples the $Y$, $J$, and $H$ bands at $0.3\arcsec$ per detector pixel \citep{Jahnke2025NISP}. We used the background-subtracted MER mosaics to inspect the candidate systems and fitted their light distributions using the associated root-mean-square (RMS) noise maps and local point-spread functions (PSFs).

We refer to the Wide Survey products tagged \texttt{DR1\_R2} as Wide R2. Three tables in the ESA Euclid Archive\footnote{\url{https://eas.esac.esa.int/}} provide the source measurements for this release. The QSO physical-parameter table gives the corrected total fluxes and their uncertainties in all four bands, together with QSO-template photometric redshift (photo-$z$) estimates. The MER catalogue provides sky positions, morphology measurements, and detection flags \citep{Euclid2026MER}. The photometric-redshift (PHZ) classification table gives the stellar probability $\pstar$ and the threshold for stellar classification.

We constructed the \emph{Wide QSO parent catalogue} by joining these tables on \texttt{object\_id} and \texttt{data\_set\_release} and selecting point sources with low stellar probabilities following \citet{Hartley2026DR1}. We required \texttt{mumax\_minus\_mag}$<-2.6$ in MER and $\pstar<$\texttt{star\_threshold} in PHZ, with finite values for all three quantities. These cuts yielded 57\,424\,066 photometric quasar candidates. For the neighbour search we further required a VIS detection (\texttt{vis\_det}$=1$), \texttt{spurious\_flag}$=0$, and a signal-to-noise ratio (S/N) of $F_b/\sigma_b\geq5$ in each band, where $F_b$ is the finite corrected total flux and $\sigma_b$ its finite positive uncertainty. This four-band subsample contains 8\,344\,260 sources and forms the input to the search in Sect.~\ref{sec:selection}.

\subsection{Reference lenses}\label{sec:reference-lenses}
The GraL catalogue compiles known lensed quasars and tabulates the sky positions of their individual images, together with source redshifts and maximum image separations where available \citep{Ducourant2026}. Of the 363 distinct GraL systems in our sample, 35 lie within the \euclid\ DR1 imaging footprint. Usable four-band images are available for 29 of these systems, including 23 in Wide R2.

For each of the 23 systems, we searched the Wide QSO parent catalogue constructed in Sect.~\ref{sec:euclid-data} for sources within $0.5\arcsec$ of the quasar image positions listed by GraL. Four systems have at least two catalogue counterparts, 12 have one each, and the remaining seven have none. We refer to the four systems with at least two counterparts as our reference lenses (green frames in Fig.~\ref{fig:recovered}). We use the colour differences between images of these reference lenses to set the colour threshold for the catalogue search (Sect.~\ref{sec:results}).

\section{Methods}\label{sec:methods}
Our search for multiple quasar images starts from the Wide QSO parent catalogue of 57\,424\,066 photometric quasar candidates constructed in Sect.~\ref{sec:euclid-data}. We searched its four-band subsample of 8\,344\,260 sources for close pairs with similar colours, using the reference lenses (Sect.~\ref{sec:reference-lenses}) to set the colour threshold. We linked pairs that shared a source and visually inspected the resulting groups (Sect.~\ref{sec:selection}). Table~\ref{tab:search-yield} gives the count at each stage, from individual catalogue sources through pairs and groups to the systems selected for follow-up. Joint fits to the four-band images separated the quasar-image candidates from the surrounding galaxy light (Sect.~\ref{sec:light}). The fitted positions then constrained a lens mass model for each system (Sect.~\ref{sec:models}).

\subsection{Searching the catalogue for multiple quasar images}\label{sec:selection}
\noindent\textit{(1) Divide the sky into cells.} Following \citet{He2023}, we searched the catalogue for close pairs. To limit the number of positions compared in each search, we assigned sources to cells using their MER sky positions and the hierarchical equal-area iso-latitude pixelization scheme \citep{Gorski2005}. At $N_{\rm side}=16$, each cell covers about 13.4 square degrees, the area of a square about $3.7^\circ$ on a side.

\noindent\textit{(2) Find neighbouring pairs.} We searched for sources within $4\arcsec$ of each catalogue entry in its own cell and in all adjacent cells. Including adjacent cells allowed us to find neighbours across cell boundaries. We retained every pair with an angular separation $0.1\arcsec<\theta<4\arcsec$, counting A--B and B--A only once. This yielded 736\,266 distinct pairs. We grouped the surviving pairs in step~(5), after the photometric cuts.

\noindent\textit{(3) Select pairs by colour and stellar probability.} The four bands give six colours per source ($\mathrm{VIS}-Y$, $\mathrm{VIS}-J$, $\mathrm{VIS}-H$, $Y-J$, $Y-H$, and $J-H$), hence 12 colour values for a pair.

For each of the six colours, we took the absolute difference between the values for A and B and divided it by two. For example, $(\mathrm{VIS}-Y)_A=0.10$~mag and $(\mathrm{VIS}-Y)_B=0.20$~mag give $|0.10-0.20|/2=0.05$~mag. We averaged the six resulting values to compute the pair score
\begin{equation}
 \scor=\max\!\left(0,\;1-\frac{1}{6}\sum_{i=1}^{6}
       \frac{|C_{A,i}-C_{B,i}|}{2\,{\rm mag}}\right),
 \label{eq:pair-colour}
\end{equation}
where $C_{A,i}$ and $C_{B,i}$ are the values of the $i$th colour for A and B. Identical colours give $\scor=1$ regardless of relative brightness. We retained 8990 pairs with $\scor\geq S_{\rm threshold}$ (set by the reference lenses in Sect.~\ref{sec:results}) and stellar probability $\pstar<0.01$ for both sources.

\noindent\textit{(4) Require reliable photometry in both sources.} We required S/N$\geq20$ in each of the four bands for each source. A pair entered the next step only if all eight flux measurements met this cut, leaving 721 pairs.

\noindent\textit{(5) Link pairs into groups.} We linked pairs sharing a source, so A--B and B--C formed one group even if A--C failed an earlier cut. An isolated pair formed its own group. The pair with the highest $\scor$ represents each group and sets its reported colour score (Table~\ref{tab:candidates}). This step retained all 721 pairs in 698 groups.

\noindent\textit{(6) Inspect the image configurations.} Of the 698 groups, 693 had usable four-band cutouts (Table~\ref{tab:search-yield}). For each of these groups we examined a $20\arcsec$ colour composite ($H$, $(Y+J)/2$, and VIS in red, green, and blue) and a $10\arcsec$ VIS image with the catalogue positions marked. For a double we required two compact images with extended light at a plausible deflector position, typically between them. A catalogue pair could also enter the sample as a candidate quad if additional compact peaks in the images outlined a plausible four-image configuration around a common centre, even when no deflector was visible. The search recovered all four reference lenses and yielded 26 additional candidates.

\begin{table}[tb]
\caption{Selection of candidate systems from the \euclid\ Wide catalogue.}\label{tab:search-yield}
\centering
\begin{tabular}{@{}lrl@{}}
\hline\hline
Selection stage & Number & Unit \\
\hline
Photometric quasar parent sample & 57\,424\,066 & sources \\
Detection flags and four-band S/N$\geq5$ & 8\,344\,260 & sources \\
$0.1\arcsec<\theta<4\arcsec$ & 736\,266 & pairs \\
Colour cut and both $\pstar<0.01$ & 8990 & pairs \\
Both sources at S/N$\geq20$ in all bands & 721 & pairs \\
Link pairs sharing a source & 698 & groups \\
Usable four-band cutouts & 693 & groups \\
Recovered reference lenses & 4 & systems \\
Additional candidates & 26 & systems \\
\hline
\end{tabular}
\end{table}

\subsection{Multiband light modelling}\label{sec:light}
We fitted the VIS, $Y$, $J$, and $H$ images jointly in $10\arcsec\times10\arcsec$ cutouts on their native mosaic grids. In each band we selected the MER PSF nearest the marked galaxy centre (24 systems), the midpoint of the two marked galaxy centres in \mbox{J0401$-$4409}, or the mean sky position of the four marked point sources in \mbox{J0239$-$3455}. We fitted a scaled copy of this PSF to each quasar image. For each galaxy we integrated an elliptical S\'ersic profile \citep{Sersic1963} over the pixels and convolved it with the same PSF.

Each point source $i$ had a single fitted sky position $\boldsymbol{\theta}^{\rm obs}_i$ across the four bands. For each galaxy $j$, the bands shared its centroid $\boldsymbol{\theta}_{{\rm G},j}$, axis ratio $q_{\rm light}$, position angle, and S\'ersic index $n$. We centred the positional prior on the MER position for each image with a distinct counterpart in the Wide QSO parent catalogue. For putative quasar images without a distinct counterpart, we marked the positions by hand on the colour composites. Each point source could move by up to $0.1\arcsec$ from its catalogue or marked position. Each fitted galaxy centroid could move by up to $0.5\arcsec$ from its own marked centre. We adopted $0.15\leq q_{\rm light}\leq1$ and $0.5\leq n\leq8$.

We fitted each galaxy's semi-major half-light radius $R_{{\rm e},b}$ independently in each band $b$, allowing $0.25\arcsec\leq R_{{\rm e},b}\leq5.5\arcsec$. At each trial geometry we also fitted, separately in each band, a non-negative flux for every point source and galaxy and a planar sky background.

We evaluated the joint fit from the pixel residuals across all four bands. We minimised
\[
  \chi^2_{\rm light}=\sum_b\sum_{i\in V_b}
  \left(\frac{D_{bi}-M_{bi}}{\sigma_{bi}}\right)^2 ,
\]
where $D_{bi}$ is the observed pixel value, $M_{bi}$ is the full model including the sky, and $\sigma_{bi}$ is the pixel uncertainty from the MER RMS map. The sum in each band runs over finite, unmasked pixels $V_b$ with positive inverse-variance weight. For \mbox{J0321$-$5722}, we applied the same sky mask in all four bands. Table~\ref{tab:candidates} gives $\chi^2_{\rm light}/\nu_{\rm light}$ for $\nu_{\rm light}=N_{\rm pix}-N_{\rm par}$. We counted fitted pixels across all bands and each free parameter once. Correlated mosaic noise makes these degrees of freedom nominal, and PSF uncertainties can contribute to the residuals.

\begin{figure*}[!t]
\centering
\includegraphics[width=\textwidth]{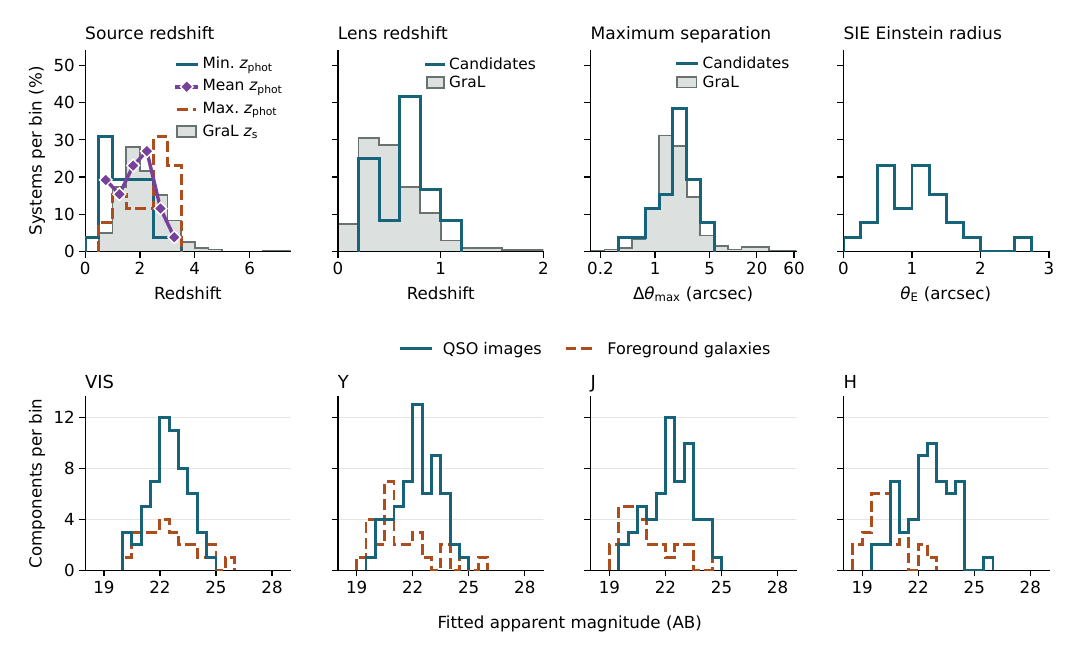}
\caption{Properties of the 26 candidates. The upper panels show the minimum, mean (purple diamonds), and maximum image photo-$z$ per system, followed by deflector redshift, maximum image separation, and SIE Einstein radius. Each distribution is normalised to the number of systems with measurements. Grey histograms show GraL lenses \citep{Ducourant2026}. The lower panels count quasar images and foreground galaxies by fitted magnitude.}
\label{fig:distributions}
\end{figure*}

\subsection{Lens mass modelling}\label{sec:models}
We fitted a single SIE \citep{Kormann1994} to the image positions $\boldsymbol{\theta}^{\rm obs}_i$ measured in Sect.~\ref{sec:light} to test whether a common source could reproduce the observed image configuration \citep{Chan2015,He2023}. Using \texttt{lenstronomy} \citep{Birrer2018,Birrer2021_lenstronomy}, we fitted the source-plane position $\boldsymbol{\beta}$, the mass centre $\boldsymbol{\theta}_{\rm mass}$, the Einstein radius $\theta_{\rm E}$, the projected axis ratio $q_{\rm mass}$, and the mass position angle. We fixed the external shear to zero and adopted $0.15\leq q_{\rm mass}\leq1$ and $0.1\,\Delta\theta_{\rm max}\leq\theta_{\rm E}\leq2\,\Delta\theta_{\rm max}$, where $\Delta\theta_{\rm max}$ is the largest observed image separation. At each trial we solved the lens equation, matched predicted and observed images one to one, and computed the positional term $\chi^2_{\rm pos}=\sum_i |\boldsymbol{\theta}^{\rm pred}_i-\boldsymbol{\theta}^{\rm obs}_i|^2/\sigma_{\rm ast}^2$, adopting $\sigma_{\rm ast}=0.05\arcsec$ per coordinate.

For systems with a measured optical centre $\boldsymbol{\theta}_{\rm opt}$, we initialised $\boldsymbol{\theta}_{\rm mass}$ there and allowed it to move within $1\arcsec$ of that position. For a single fitted galaxy, $\boldsymbol{\theta}_{\rm opt}$ was its S\'ersic centroid $\boldsymbol{\theta}_{{\rm G},1}$. For the two galaxies in \mbox{J0401$-$4409}, we took the sky midpoint of $\boldsymbol{\theta}_{{\rm G},1}$ and $\boldsymbol{\theta}_{{\rm G},2}$ as $\boldsymbol{\theta}_{\rm opt}$ for the single SIE. For \mbox{J0239$-$3455}, which has no measured optical centre, we started at the mean sky position of its four images and used a prior radius of $3\arcsec$ around that position.

We used different constraints for doubles and quads because their image positions supply four and eight coordinates, respectively, for seven free parameters. For doubles we used the VIS flux ratio as a weak prior on the magnification ratio. Stellar microlensing can alter the observed continuum flux ratio \citep{Motta2012,Zheng2022}, so we downweighted its contribution and minimised
\begin{equation}
 Q=\chi^2_{\rm pos}+0.01\left(\frac{|\mu_B/\mu_A|-F_B/F_A}{\sigma_{B/A}}\right)^2 ,
\end{equation}
where $F_B/F_A$ is the fitted VIS flux ratio, $|\mu_B/\mu_A|$ is the predicted magnification ratio, and $\sigma_{B/A}$ is the uncertainty propagated from the two conditional flux errors assuming independence. For quads the eight coordinates give one nominal degree of freedom, so we fitted the positions alone ($Q=\chi^2_{\rm pos}$). We tried multiple starting configurations and checked the predicted positions with an independent angular root search. For effectively equal $Q$ we favoured solutions that reproduced the observed multiplicity and, among converged solutions, those closest to the adopted centre.

We evaluated each fit by the two-dimensional image-plane RMS, $R_{\rm img}=[N_{\rm img}^{-1}\sum_i|\boldsymbol{\theta}^{\rm pred}_i-\boldsymbol{\theta}^{\rm obs}_i|^2]^{1/2}$, where $N_{\rm img}$ is the number of matched images. For the 25 systems with an optical centre we also measured $\Delta_{\rm LM}$, the angular separation between $\boldsymbol{\theta}_{\rm opt}$ and $\boldsymbol{\theta}_{\rm mass}$. For \mbox{J0239$-$3455}, which has no measured optical centre, we report only $R_{\rm img}$ (Table~\ref{tab:candidates}).

\section{Results}\label{sec:results}

Our search recovered all four reference lenses, the known systems with at least two quasar images in the Wide QSO parent catalogue (Sect.~\ref{sec:reference-lenses}). Their image pairs have colour scores from 0.881 to 0.986. We adopted $\scor\geq0.85$ to allow slightly larger colour differences than those in the reference pairs, corresponding to a mean absolute difference of at most 0.30~mag across the six colours (Eq.~\ref{eq:pair-colour}).

The catalogue search and visual inspection yielded 26 additional candidates, with 23 double-image and three four-image configurations (Table~\ref{tab:search-yield}). We show their population properties in Fig.~\ref{fig:distributions} and the light fits for four representative systems in Fig.~\ref{fig:examples}. Table~\ref{tab:candidates} gives the candidate positions, image and deflector photo-$z$ estimates, and SIE Einstein radii. It also reports the reduced $\chi^2$ of the light fits and the image-position RMS and light--mass offset of the SIE fits. Table~\ref{tab:component-photometry} lists the four-band magnitudes of the individual images and foreground galaxies.

The candidate source redshifts broadly overlap the main GraL distribution (first panel of Fig.~\ref{fig:distributions}). The Wide QSO parent catalogue provides photo-$z$ estimates for all 53 catalogue-selected images. We marked five additional putative images on the colour composites and found catalogue redshifts for two of them within $0.5\arcsec$ of their fitted positions. For each system we show the minimum, arithmetic mean, and maximum of the available image redshifts. The median of the 26 system means is 1.81, compared with a median source redshift of 2.00 for the 347 GraL lenses with a measurement.

The median deflector redshift is 0.67 for the 12 candidate galaxies with photo-$z$ estimates and 0.44 for the 203 GraL deflectors with measured redshifts (second panel of Fig.~\ref{fig:distributions}). We identified the 12 galaxies by matching the 26 fitted galaxy centres to Wide R2 MER sources within $0.5\arcsec$, excluding entries assigned to a quasar image or within $0.5\arcsec$ of a fitted image. Their PHZ redshifts span 0.33--1.12 (Table~\ref{tab:candidates}). Nine of the 12 lie above the GraL median, adding candidates on the higher-redshift side of the known deflector distribution.

\begin{table*}[!t]
\centering
\caption{Catalogue properties, image and foreground-galaxy photo-$z$ estimates, and light and SIE fits for the 26 candidates.}\label{tab:candidates}
\small
\setlength{\tabcolsep}{1.5pt}
\begin{tabular*}{\textwidth}{@{\extracolsep{\fill}}l*{16}{r}@{}}
\hline\hline
 & \multicolumn{3}{c}{Representative pair} & & \multicolumn{4}{c}{Image photo-$z$} & Galaxy & \multicolumn{3}{c}{Light fit} & \multicolumn{4}{c}{SIE fit} \\
\cline{2-4}\cline{6-9}\cline{11-13}\cline{14-17}
System & RA & Dec & $\scor$ & $\mathrm{Sep}_{\max}$ & A & B & C & D & $z_{\rm phot,G}$ & $N_{\rm img}$ & $N_{\rm Ser}$ & $\chi^2_{\nu,{\rm light}}$ & $\theta_{\rm E}$ & $q_{\rm mass}$ & $R_{\rm img}$ & $\Delta_{\rm LM}$ \\
 & (deg) & (deg) &  & (\arcsec) &  &  &  &  &  &  &  &  & (\arcsec) &  & (\arcsec) & (\arcsec) \\
\hline
\mbox{J0238$-$3456} & $39.72511$ & $-34.93840$ & $0.967$ & $1.01$ & $0.95$ & $0.80$ &  &  & $\ldots$ & $2$ & $1$ & $0.40$ & $0.519$ & $0.85$ & $<0.001$ & $<0.001$ \\
\mbox{J0239$-$3455} & $39.78395$ & $-34.93244$ & $0.987$ & $1.03$ & $2.60$ & $1.30$ & $\ldots$ & $\ldots$ & $\ldots$ & $4$ & $0$ & $0.93$ & $0.549$ & $0.95$ & $0.00063$ & $\ldots$ \\
\mbox{J0321$-$5722} & $50.49858$ & $-57.37370$ & $0.916$ & $2.41$ & $2.15$ & $0.70$ &  &  & $\ldots$ & $2$ & $1$ & $0.20$ & $1.200$ & $0.96$ & $<0.001$ & $0.036$ \\
\mbox{J0323$-$4222} & $50.78933$ & $-42.36763$ & $0.993$ & $0.56$ & $0.25$ & $1.55$ &  &  & $\ldots$ & $2$ & $1$ & $0.24$ & $0.294$ & $0.76$ & $<0.001$ & $0.218$ \\
\mbox{J0401$-$4409} & $60.36177$ & $-44.15667$ & $0.932$ & $1.94$ & $2.00$ & $3.15$ &  &  & $\ldots$ & $2$ & $2$ & $0.31$ & $1.020$ & $0.86$ & $<0.001$ & $0.043$ \\
\mbox{J0432$-$6627} & $68.06443$ & $-66.46195$ & $0.888$ & $1.76$ & $1.40$ & $1.95$ &  &  & $\ldots$ & $2$ & $1$ & $2.45$ & $0.933$ & $0.63$ & $<0.001$ & $<0.001$ \\
\mbox{J0436$-$6418} & $69.14253$ & $-64.30802$ & $0.925$ & $2.76$ & $0.75$ & $3.00$ &  &  & $0.70$ & $2$ & $1$ & $0.55$ & $1.423$ & $0.88$ & $<0.001$ & $<0.001$ \\
\mbox{J0437$-$6505} & $69.35944$ & $-65.09889$ & $0.910$ & $2.89$ & $2.65$ & $1.50$ &  &  & $0.33$ & $2$ & $1$ & $0.30$ & $1.439$ & $0.96$ & $<0.001$ & $0.111$ \\
\mbox{J0441$-$6401} & $70.41338$ & $-64.02339$ & $0.946$ & $5.06$ & $2.90$ & $2.95$ & $2.55$ & $1.55$ & $0.36$ & $4$ & $1$ & $1.04$ & $2.637$ & $0.79$ & $0.206$ & $0.703$ \\
\mbox{J0442$-$6523} & $70.60181$ & $-65.38594$ & $0.966$ & $2.47$ & $1.50$ & $1.75$ &  &  & $0.33$ & $2$ & $1$ & $16.82$ & $1.287$ & $0.56$ & $<0.001$ & $0.141$ \\
\mbox{J0444$-$6415} & $71.00839$ & $-64.25367$ & $0.986$ & $2.67$ & $2.55$ & $2.05$ &  &  & $0.93$ & $2$ & $1$ & $0.99$ & $1.241$ & $0.45$ & $<0.001$ & $0.494$ \\
\mbox{J0445$-$6424} & $71.36429$ & $-64.41002$ & $0.928$ & $1.98$ & $2.90$ & $3.00$ &  &  & $0.61$ & $2$ & $1$ & $0.35$ & $1.015$ & $0.96$ & $<0.001$ & $0.002$ \\
\mbox{J0448$-$6432} & $72.15119$ & $-64.53768$ & $0.870$ & $2.49$ & $1.90$ & $2.15$ &  &  & $0.82$ & $2$ & $1$ & $1.31$ & $1.375$ & $0.96$ & $<0.001$ & $0.048$ \\
\mbox{J0448$-$6433} & $72.07309$ & $-64.56262$ & $0.935$ & $2.97$ & $2.05$ & $2.50$ &  &  & $0.73$ & $2$ & $1$ & $0.29$ & $1.507$ & $0.98$ & $<0.001$ & $0.082$ \\
\mbox{J0449$-$6436} & $72.34065$ & $-64.60126$ & $0.913$ & $2.09$ & $1.00$ & $0.85$ &  &  & $\ldots$ & $2$ & $1$ & $0.50$ & $1.148$ & $0.87$ & $<0.001$ & $0.005$ \\
\mbox{J0512$-$2438} & $78.04921$ & $-24.64727$ & $0.973$ & $1.18$ & $1.40$ & $1.40$ &  &  & $\ldots$ & $2$ & $1$ & $0.26$ & $0.578$ & $0.79$ & $<0.001$ & $0.040$ \\
\mbox{J0525$-$5955} & $81.39167$ & $-59.92428$ & $0.962$ & $1.50$ & $2.50$ & $1.00$ &  &  & $\ldots$ & $2$ & $1$ & $1.32$ & $0.711$ & $0.45$ & $<0.001$ & $0.066$ \\
\mbox{J0526$-$6112} & $81.53507$ & $-61.20385$ & $0.985$ & $1.18$ & $3.40$ & $3.45$ &  &  & $\ldots$ & $2$ & $1$ & $0.27$ & $0.610$ & $0.45$ & $<0.001$ & $0.026$ \\
\mbox{J0527$-$6132} & $81.90325$ & $-61.53984$ & $0.921$ & $0.84$ & $1.25$ & $1.30$ &  &  & $\ldots$ & $2$ & $1$ & $1.34$ & $0.432$ & $0.89$ & $<0.001$ & $0.460$ \\
\mbox{J0527$-$6136} & $81.82798$ & $-61.61532$ & $0.868$ & $2.01$ & $1.10$ & $0.75$ &  &  & $\ldots$ & $2$ & $1$ & $0.25$ & $1.010$ & $0.77$ & $<0.001$ & $<0.001$ \\
\mbox{J0539$-$6051} & $84.85915$ & $-60.86413$ & $0.930$ & $1.73$ & $2.85$ & $2.25$ &  &  & $0.65$ & $2$ & $1$ & $0.34$ & $0.920$ & $0.55$ & $<0.001$ & $0.040$ \\
\mbox{J0546$-$6132} & $86.72043$ & $-61.53754$ & $0.937$ & $3.28$ & $0.80$ & $0.80$ &  &  & $1.12$ & $2$ & $1$ & $0.55$ & $1.709$ & $0.92$ & $<0.001$ & $<0.001$ \\
\mbox{J0548$-$6116} & $87.17686$ & $-61.26827$ & $0.979$ & $1.77$ & $0.70$ & $2.15$ &  &  & $0.68$ & $2$ & $1$ & $0.17$ & $0.923$ & $0.76$ & $<0.001$ & $0.060$ \\
\mbox{J0600$-$4356} & $90.09344$ & $-43.93490$ & $0.940$ & $3.87$ & $1.50$ & $3.45$ &  &  & $0.53$ & $2$ & $1$ & $0.38$ & $1.966$ & $0.90$ & $<0.001$ & $0.022$ \\
\mbox{J0602$-$3852} & $90.54824$ & $-38.87034$ & $0.979$ & $1.34$ & $0.85$ & $0.85$ & $\ldots$ & $3.20$ & $\ldots$ & $4$ & $1$ & $4.48$ & $0.685$ & $0.79$ & $0.0022$ & $0.024$ \\
\mbox{J1808$+$7607} & $272.15167$ & $+76.13165$ & $0.973$ & $0.42$ & $2.25$ & $2.55$ &  &  & $\ldots$ & $2$ & $1$ & $0.19$ & $0.210$ & $1.00$ & $<0.001$ & $0.380$ \\
\hline
\end{tabular*}

\tablefoot{Right ascension (RA) and declination (Dec) give the representative pair midpoint in the International Celestial Reference System. $\scor$ is the pair colour score and $\mathrm{Sep}_{\max}$ the largest fitted image separation. $z_{\rm phot,G}$ is the PHZ posterior median for the galaxy. Blank C and D cells denote doubles. Dots mark unavailable redshifts. $N_{\rm img}$ and $N_{\rm Ser}$ count fitted images and galaxies. $R_{\rm img}$ is the image-position RMS and $\Delta_{\rm LM}$ the light--mass offset (Sect.~\ref{sec:models}).}
\end{table*}

The candidates have maximum image separations comparable to those of known GraL lenses (third panel of Fig.~\ref{fig:distributions}). The largest angular separation between fitted images, $\Delta\theta_{\rm max}$, ranges from $0.42\arcsec$ to $5.06\arcsec$ across the 26 candidates. Its median is $1.96\arcsec$, compared with $1.72\arcsec$ for the 347 GraL lenses with a measured separation. The SIE fits give Einstein radii $\theta_{\rm E}$ from $0.21\arcsec$ to $2.64\arcsec$, with a median of $1.01\arcsec$ (fourth panel).

On the AB magnitude system, the quasar images have median magnitudes close to 22.5 in all four bands, while the foreground galaxies brighten towards the near-infrared (lower panels of Fig.~\ref{fig:distributions}). In VIS, $Y$, $J$, and $H$, the median image magnitudes are 22.53, 22.32, 22.31, and 22.55, respectively. The corresponding foreground-galaxy medians are 22.41, 20.90, 20.54, and 20.12.

\section{Discussion and conclusions}\label{sec:discussion}
Only two of the 23 known lensed quasars in our \euclid\ Wide R2 sample show conspicuous arcs in VIS (Fig.~\ref{fig:recovered}). This motivated a search based on the multiple compact images of the quasar. Following the approach developed for the DESI Legacy Imaging Surveys \citep{He2022,He2023}, we first selected photometric quasar candidates and then searched that catalogue for close groups with similar colours.

From 698 groups selected for visual inspection, we identified 26 high-quality lensed-quasar candidates. None has a positional match within $3\arcsec$ in the ECLIPSE galaxy--galaxy lens catalogue \citep{Lines2026ECLIPSE}, illustrating how searches for compact quasar images complement those for extended arcs. The main limitation of our search is that at least two images must enter the Wide QSO parent catalogue. Only four of the 23 known lenses meet this requirement, and we recovered all four, giving an overall recovery fraction of 17.4\% (Sect.~\ref{sec:reference-lenses}).

In future searches we will relax the selection criteria for the second image to improve recovery. We tested this approach on the 12 known lenses with a single counterpart in the parent catalogue (yellow frames in Fig.~\ref{fig:recovered}). For each counterpart, we searched the full \euclid\ MER catalogue for a companion with similar VIS, $Y$, $J$, and $H$ colours, relaxing the compactness cut on the companion from \texttt{mumax\_minus\_mag}$<-2.6$ to $<-1$. This recovered eight additional lenses. Searching Gaia Data Release~3 (DR3) for separately catalogued neighbours \citep{GaiaDR3overview} recovered two more, bringing the total to 14 of 23 (60.9\%). If we require the parent source in the MER companion search to have a Gaia DR3 counterpart, we estimate that about 8000 Wide R2 cutouts would need visual inspection.

The seven known lenses with no parent-catalogue entry (30\% of the sample, red frames in Fig.~\ref{fig:recovered}) require a different starting point. In five systems, MER records a single extended or blended source (red crosses). In the other two, MER detections coincide with known quasar images but fail the point-source criterion, \texttt{mumax\_minus\_mag}$<-2.6$ (yellow crosses). We could search around extended MER sources or look directly in \euclid\ images for multiple point sources and arcs. Convolutional networks have been used to classify sources in survey images \citep{Wang2022} and to identify lensed-quasar candidates in multiband imaging \citep{Andika2023,He2025-Apj}.

To summarize, our search demonstrates the potential of \euclid\ for finding lensed quasars. We identified 26 high-quality candidates while recovering 17.4\% of the known lenses in Wide R2. As an illustrative extrapolation, suppose that recovering all the known lenses would bring the candidate sample to about 126 across the present 1900~deg$^2$. At the same surface density, the full 14\,000~deg$^2$ Wide Survey would yield more than 900 candidates.

\begin{acknowledgements}
This work has made use of data from the European Space Agency (ESA) mission \euclid\ (\url{https://www.cosmos.esa.int/euclid}), processed by the Euclid Consortium (\url{https://www.euclid-ec.org/}).
We obtained the DR1-Foundation data from the ESA Euclid Archive and the Gaia GraL catalogue through the Centre de donn\'ees astronomiques de Strasbourg.
Z.H. acknowledges support from the National Natural Science Foundation of China (Grant No.~12403104) and the Ganpo High-Level Innovative Talent Program (No.~gpyc20260051).
F.S. acknowledges support from the National Natural Science Foundation of China (Grant No.~12375049) and the Ganpo High-Level Innovative Talent Program (No.~gpyc20240060).
Z.H. thanks Ran Li, Yiping Shu, Nan Li, Tian Li, and Xinzhong Er for helpful discussions.
\end{acknowledgements}
\bibliographystyle{aa}
\bibliography{cite-database}

\clearpage
\onecolumn
\appendix
\section{Candidate images and lens models}\label{app:catalogue}
\begin{figure}[!ht]
\centering
\includegraphics[page=1,width=\textwidth]{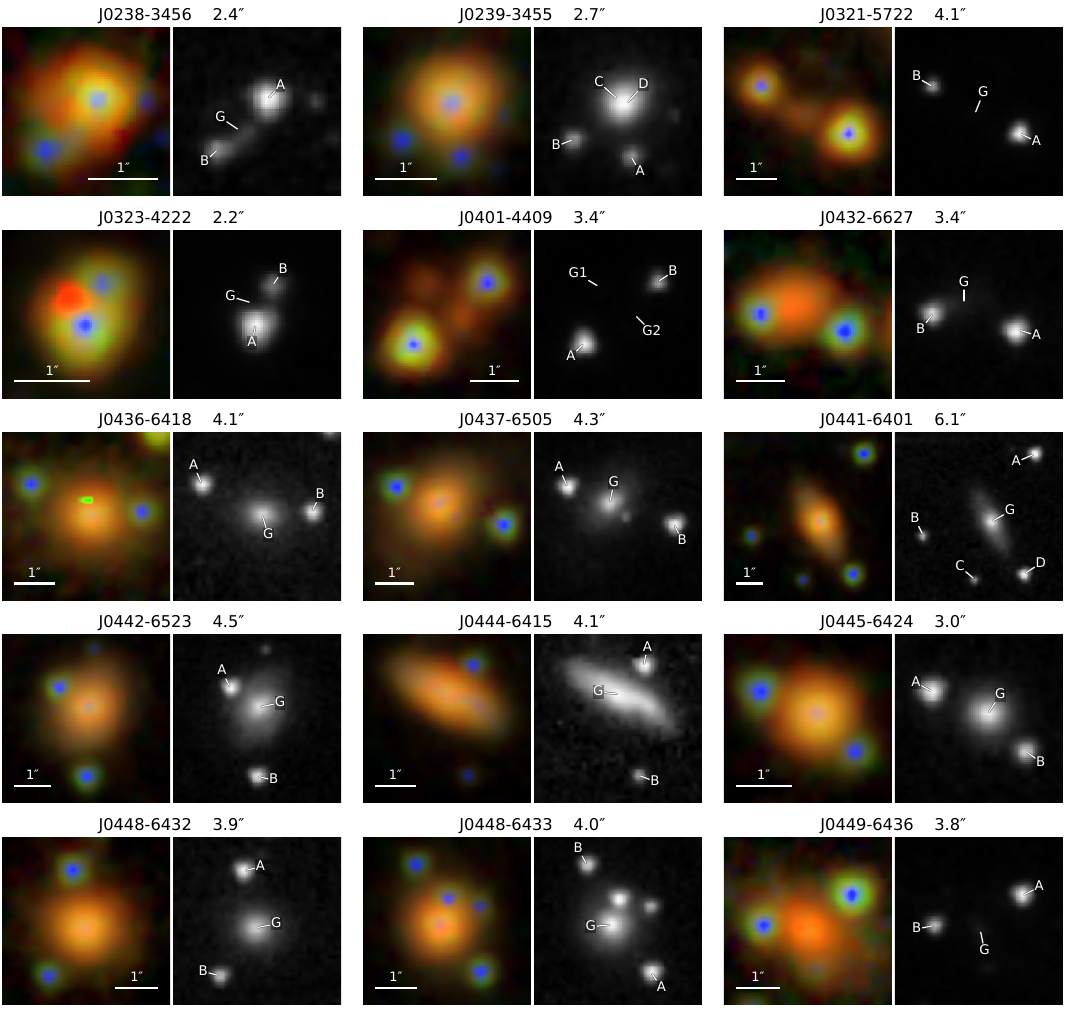}
\caption{Euclid images of the 26 shortlisted systems in the order of Table~\ref{tab:candidates}. Each system is shown in colour (left) and VIS (right), with the same centre and field of view in both panels. The composite maps $H$, $(Y+J)/2$, and VIS to red, green, and blue, respectively. The headings give the side length of each square field. The white bar marks $1\arcsec$. A--D label the fitted point sources on the VIS image. G marks the fitted galaxy centre (G1 and G2 in \mbox{J0401$-$4409}). Only point sources were fitted in \mbox{J0239$-$3455}. North is up and east is left.}
\label{fig:gallery}
\end{figure}

\begin{figure}[t]
\centering
\includegraphics[page=1,width=\textwidth]{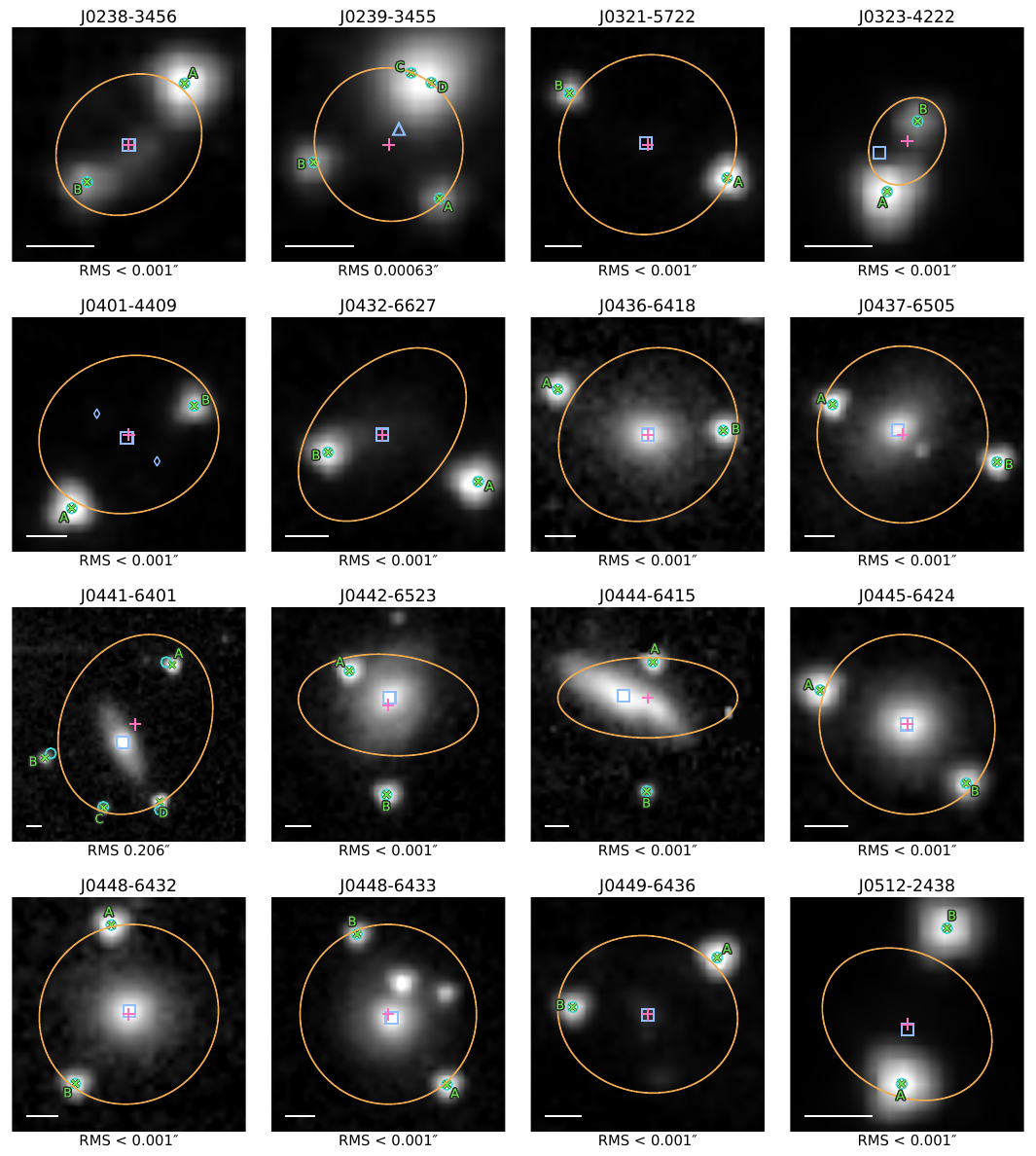}
\caption{Single-SIE fits to the 26 systems in the same reading order as Fig.~\ref{fig:gallery}. Green crosses labelled A--D mark the measured image positions on the VIS cutout. Cyan circles mark the predicted images, and the orange line is the tangential critical curve. A magenta plus marks the fitted mass centre. The blue square marks the optical reference position. In \mbox{J0401$-$4409} the two blue diamonds mark the individual galaxy light centroids, and the blue square is their equal-weight midpoint. In \mbox{J0239$-$3455} a triangle marks the mean image position used to initialise the model. Each panel gives the system name above the image and the two-dimensional image-plane RMS below it. North is up, east is left, and each white bar spans $0.5\arcsec$. Fields include the entire critical curve.}
\label{fig:sie}
\end{figure}

Figure~\ref{fig:gallery} shows the 26 candidates in the order of Table~\ref{tab:candidates}. Figure~\ref{fig:examples} shows the light fits for the four systems discussed below, and Fig.~\ref{fig:sie} shows the SIE fits for the full sample. Table~\ref{tab:component-photometry} gives their component photometry and galaxy structure.

The accompanying files \texttt{candidate\_systems.csv} and \texttt{candidate\_members.csv} give, respectively, group measurements and individual MER object identifiers, positions, VIS magnitudes, stellar probabilities and QSO photo-$z$ estimates. Object identifiers are signed 64-bit integers. The accompanying \texttt{model\_diagnostics.csv} gives the unrounded fit statistics and degrees of freedom.

The archive table names are \texttt{dr1.phz\_qso\_physical\_parameters\_wide}, \texttt{dr1.mer\_catalogue\_wide\_survey}, and \texttt{dr1.phz\_classification\_wide}. The accompanying \texttt{parent\_sample.adql} gives the archive query, including the joins on \texttt{object\_id} and \texttt{data\_set\_release} and the parent-sample cuts of Sect.~\ref{sec:euclid-data}. The images are from background-subtracted MER mosaics in the same release. For each candidate in Fig.~\ref{fig:gallery}, we used a common field for the colour composite and VIS image, enclosing all fitted point sources and the visible extent of the central galaxy light.
\clearpage
\begin{figure}[t]
\centering
\includegraphics[page=2,width=\textwidth]{figures/gallery.pdf}
\par\vspace{\VSpaceBeforeFigCaption}\noindent
\parbox{\textwidth}{\small\textbf{Fig.~\ref{fig:gallery}.} \textit{Continued.} Each pair shows the colour composite (left) and the VIS image (right). The heading gives the field width and the white bar marks $1\arcsec$.}
\end{figure}
\clearpage
\subsection*{Notes on individual systems}
\begin{figure}[!ht]
\centering
\includegraphics[width=0.83\textwidth]{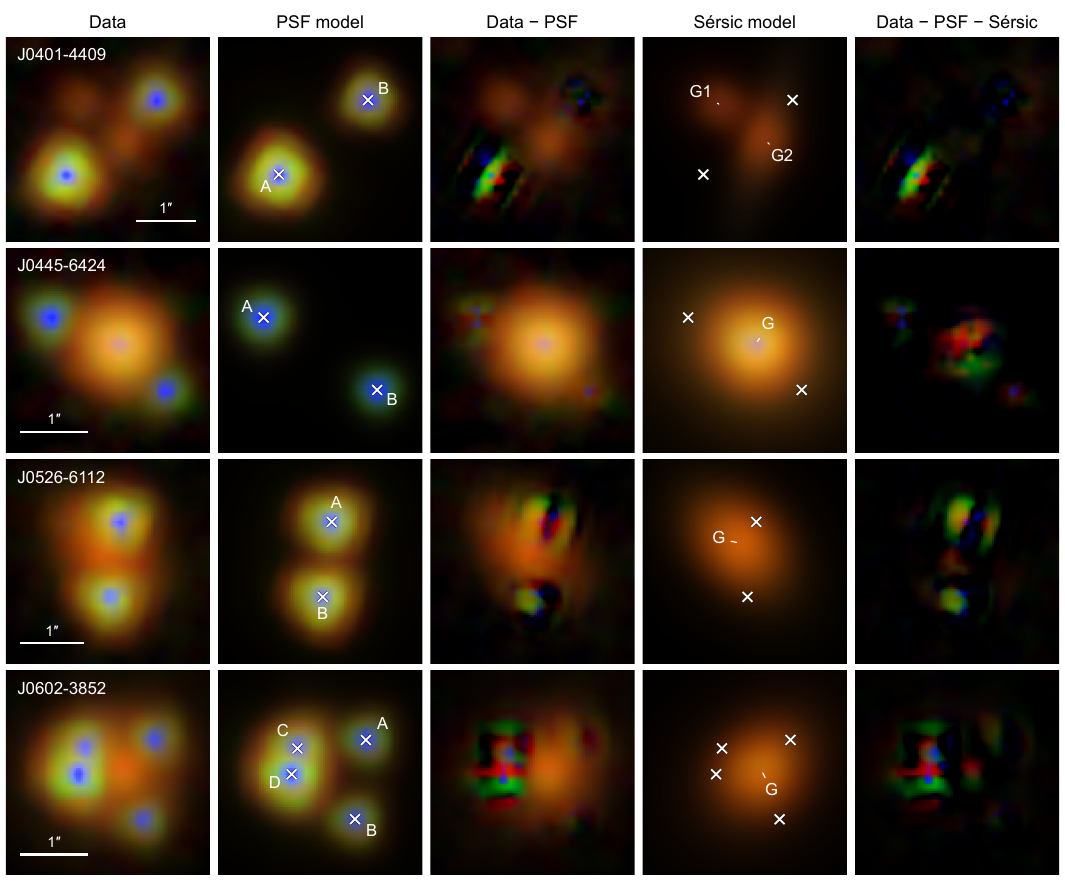}
\caption{Multiband image fits for four candidate systems. From left to right, the columns show the Euclid data, the point-source model, the data after point-source subtraction, the PSF-convolved S\'ersic model, and the residual after subtracting both models. Each row uses the same field as Fig.~\ref{fig:gallery} and a common colour stretch across its five panels ($H$, $(Y+J)/2$, and VIS in red, green, and blue). White crosses in both model panels mark the fitted point-source positions (A--D as applicable). G marks the fitted galaxy centre, with G1 and G2 denoting the two components in \mbox{J0401$-$4409}. North is up, east is left, and each bar spans $1\arcsec$.}
\label{fig:examples}
\end{figure}

\begingroup
\patchcmd{\paragraph}{-16dd}{-6dd}{}{}
\paragraph{J0401$-$4409.}
The catalogue photo-$z$ estimates for images A and B are 2.00 and 3.15, respectively. After PSF subtraction, we resolve two galaxies, G1 and G2, whose S\'ersic centroids lie $0.93\arcsec$ apart (Fig.~\ref{fig:examples}). Neither has a catalogue redshift. A single SIE with $\theta_{\rm E}=1.02\arcsec$ fits the double with its mass centre $0.043\arcsec$ from their midpoint, the centre of the prior.

\paragraph{J0445$-$6424.}
Both images have photo-$z$ estimates near three ($z_{\rm phot,A}=2.90$ and $z_{\rm phot,B}=3.00$). The bright red galaxy between them has $z_{\rm phot,G}=0.61$ (90\% interval 0.58--0.70), placing it in the foreground. The best-fitting SIE is nearly circular ($q_{\rm mass}=0.96$), with its mass centre $0.002\arcsec$ from the galaxy's light centroid.

\paragraph{J0526$-$6112.}
Both images have photo-$z$ estimates near 3.4 (3.40 for A and 3.45 for B) and similar VIS magnitudes (21.42 and 21.56). Subtracting them reveals a red galaxy ($\mathrm{VIS}-H=3.20$~mag) between the images, with no separate catalogue redshift. The fitted Einstein radius of $0.61\arcsec$ is close to half the $1.18\arcsec$ image separation.

\paragraph{J0602$-$3852.}
In this candidate quad, A and B both have $z_{\rm phot}=0.85$. Images C and D lie $0.39\arcsec$ apart and share one catalogue entry with $z_{\rm phot}=3.20$, listed under D in Table~\ref{tab:candidates}. The deflector has no catalogue redshift. Assuming a deflector redshift of $z_{\rm d}=0.50$, the projected mass within the Einstein radius ($0.685\arcsec$, or 4.18~kpc) is $1.02\times10^{11}\,M_\odot$ for $z_{\rm s}=3.20$ and $2.05\times10^{11}\,M_\odot$ for $z_{\rm s}=0.85$.\footnote{We compute the projected Einstein mass as $M_{\rm E}=(c^2/4G)\,\theta_{\rm E}^2 D_{\rm d}D_{\rm s}/D_{\rm ds}$ using angular diameter distances in a flat $\Lambda$CDM cosmology with $H_0=70$~km~s$^{-1}$~Mpc$^{-1}$ and $\Omega_{\rm m}=0.3$.} The position-only SIE ($R_{\rm img}=0.0022\arcsec$) predicts $|\mu_D/\mu_C|=0.88$, while the light fits give $F_D/F_C$ of 2.59, 2.21, 2.91, and 2.06 in VIS, $Y$, $J$, and $H$, respectively. This reversed brightness ordering suggests a flux-ratio anomaly, although structured residuals around C and D make the measured ratio sensitive to deblending.
\endgroup

\clearpage
\clearpage
\begin{figure}[t]
\centering
\includegraphics[page=2,width=\textwidth]{figures/sie_models.pdf}
\par\vspace{\VSpaceBeforeFigCaption}\noindent
\parbox{\textwidth}{\small\textbf{Fig.~\ref{fig:sie}.} \textit{Continued.} Symbols are as on the preceding page. Each white bar spans $0.5\arcsec$.}
\end{figure}
\clearpage
\begingroup
\normalsize
\setlength{\tabcolsep}{7pt}
\renewcommand{\arraystretch}{1.05}
\setlength{\LTcapwidth}{\textwidth}
\newcommand{\componenthead}{%
\hline\hline
 & & \multicolumn{4}{c}{Apparent magnitude (AB)} & \multicolumn{4}{c}{Half-light radius $R_{\rm e}$ (\arcsec)} & Axis ratio \\
\cline{3-6}\cline{7-10}
System & Component & VIS & $Y$ & $J$ & $H$ & VIS & $Y$ & $J$ & $H$ & $q_{\rm light}$ \\
\hline}
\begin{longtable}{@{}ll*{4}{r}@{\hspace{1.25em}}*{4}{r}@{\hspace{1.25em}}r@{}}
\caption{Four-band photometry of the fitted quasar images and foreground galaxies, and structural parameters of the galaxies.}\label{tab:component-photometry}\\
\componenthead
\endfirsthead
\caption[]{\textit{Continued.}}\\
\componenthead
\endhead
\hline
\endfoot
\mbox{J0238$-$3456} & A & 22.34 & 21.34 & 21.04 & 20.87 &  &  &  &  &  \\*
 & B & 24.17 & 24.27 & 24.46 & 25.61 &  &  &  &  &  \\*
 & G & 24.85 & 23.68 & 22.76 & 22.40 & 0.25 & 0.27 & 0.40 & 0.47 & 0.56 \\[3pt]
\mbox{J0239$-$3455} & A & 23.99 & 23.83 & 23.76 & 23.84 &  &  &  &  &  \\*
 & B & 23.84 & 23.84 & 23.85 & 24.04 &  &  &  &  &  \\*
 & C & 22.60 & 21.52 & 21.33 & 21.19 &  &  &  &  &  \\*
 & D & 22.20 & 21.59 & 21.01 & 20.73 &  &  &  &  &  \\[3pt]
\mbox{J0321$-$5722} & A & 21.34 & 20.88 & 20.75 & 20.62 &  &  &  &  &  \\*
 & B & 22.74 & 22.34 & 22.03 & 21.76 &  &  &  &  &  \\*
 & G & 22.41 & 21.45 & 21.31 & 21.01 & 5.50 & 5.50 & 2.89 & 2.61 & 0.66 \\[3pt]
\mbox{J0323$-$4222} & A & 20.39 & 19.92 & 19.82 & 19.85 &  &  &  &  &  \\*
 & B & 22.45 & 21.62 & 21.61 & 22.34 &  &  &  &  &  \\*
 & G & $\ldots$ & 23.77 & 22.72 & 20.65 & $\ldots$ & 0.25 & 0.25 & 0.25 & 0.21 \\[3pt]
\mbox{J0401$-$4409} & A & 20.94 & 20.48 & 20.39 & 20.37 &  &  &  &  &  \\*
 & B & 22.24 & 22.09 & 22.06 & 21.86 &  &  &  &  &  \\*
 & G1 & 25.72 & 25.61 & 23.43 & 22.68 & 0.33 & 0.25 & 0.25 & 0.34 & 0.51 \\*
 & G2 & 23.11 & 22.80 & 22.45 & 22.01 & 5.50 & 2.51 & 0.73 & 0.58 & 0.15 \\[3pt]
\mbox{J0432$-$6627} & A & 22.46 & 22.38 & 22.53 & 22.71 &  &  &  &  &  \\*
 & B & 22.82 & 22.77 & 22.91 & 23.04 &  &  &  &  &  \\*
 & G & 24.03 & 22.43 & 21.89 & 21.32 & 0.45 & 0.40 & 0.34 & 0.28 & 0.76 \\[3pt]
\mbox{J0436$-$6418} & A & 23.24 & 23.15 & 23.35 & 23.59 &  &  &  &  &  \\*
 & B & 23.38 & 23.29 & 23.44 & 23.57 &  &  &  &  &  \\*
 & G & 21.36 & 20.01 & 19.84 & 19.57 & 1.59 & 1.51 & 1.14 & 0.91 & 0.74 \\[3pt]
\mbox{J0437$-$6505} & A & 21.90 & 22.04 & 22.12 & 22.54 &  &  &  &  &  \\*
 & B & 21.92 & 21.91 & 22.07 & 22.20 &  &  &  &  &  \\*
 & G & 20.31 & 19.40 & 19.11 & 18.86 & 1.15 & 0.96 & 0.78 & 0.60 & 0.62 \\[3pt]
\mbox{J0441$-$6401} & A & 22.16 & 22.21 & 22.27 & 22.56 &  &  &  &  &  \\*
 & B & 23.61 & 23.39 & 23.32 & 23.50 &  &  &  &  &  \\*
 & C & 24.18 & 23.95 & 24.21 & 24.22 &  &  &  &  &  \\*
 & D & 22.34 & 22.41 & 22.40 & 22.66 &  &  &  &  &  \\*
 & G & 20.97 & 19.91 & 19.58 & 19.26 & 0.46 & 0.39 & 0.35 & 0.33 & 0.35 \\[3pt]
\mbox{J0442$-$6523} & A & 22.23 & 22.28 & 22.35 & 22.54 &  &  &  &  &  \\*
 & B & 22.59 & 22.59 & 22.65 & 22.93 &  &  &  &  &  \\*
 & G & 20.76 & 19.94 & 19.58 & 19.29 & 0.44 & 0.41 & 0.39 & 0.36 & 0.77 \\[3pt]
\mbox{J0444$-$6415} & A & 22.88 & 22.84 & 22.97 & 23.02 &  &  &  &  &  \\*
 & B & 24.56 & 24.68 & 24.51 & 24.34 &  &  &  &  &  \\*
 & G & 20.55 & 19.63 & 19.23 & 18.89 & 0.72 & 0.66 & 0.62 & 0.57 & 0.27 \\[3pt]
\mbox{J0445$-$6424} & A & 22.93 & 23.04 & 23.23 & 23.49 &  &  &  &  &  \\*
 & B & 23.62 & 23.68 & 23.86 & 24.40 &  &  &  &  &  \\*
 & G & 21.84 & 20.70 & 20.34 & 20.00 & 0.25 & 0.25 & 0.25 & 0.25 & 0.82 \\[3pt]
\mbox{J0448$-$6432} & A & 23.04 & 23.05 & 23.07 & 23.12 &  &  &  &  &  \\*
 & B & 23.87 & 23.69 & 23.93 & 24.17 &  &  &  &  &  \\*
 & G & 22.07 & 20.72 & 20.33 & 19.95 & 0.29 & 0.29 & 0.27 & 0.25 & 0.86 \\[3pt]
\pagebreak
\mbox{J0448$-$6433} & A & 22.66 & 22.59 & 22.72 & 22.96 &  &  &  &  &  \\*
 & B & 23.21 & 23.16 & 23.43 & 23.58 &  &  &  &  &  \\*
 & G & 21.01 & 19.90 & 19.61 & 19.20 & 0.35 & 0.30 & 0.29 & 0.25 & 0.89 \\[3pt]
\mbox{J0449$-$6436} & A & 21.43 & 21.39 & 21.51 & 21.69 &  &  &  &  &  \\*
 & B & 22.39 & 22.44 & 22.53 & 22.66 &  &  &  &  &  \\*
 & G & 22.18 & 20.88 & 20.53 & 20.12 & 2.12 & 1.19 & 0.81 & 0.62 & 0.56 \\[3pt]
\mbox{J0512$-$2438} & A & 20.66 & 20.41 & 20.44 & 21.09 &  &  &  &  &  \\*
 & B & 21.03 & 20.69 & 20.62 & 20.81 &  &  &  &  &  \\*
 & G & 24.54 & 24.62 & 24.56 & 20.35 & 5.50 & 0.57 & 0.25 & 0.25 & 0.15 \\[3pt]
\mbox{J0525$-$5955} & A & 22.11 & 22.21 & 22.37 & 22.64 &  &  &  &  &  \\*
 & B & 23.94 & 23.94 & 24.15 & 24.37 &  &  &  &  &  \\*
 & G & 22.73 & 20.76 & 20.46 & 19.96 & 5.50 & 5.43 & 4.60 & 5.26 & 0.28 \\[3pt]
\mbox{J0526$-$6112} & A & 21.42 & 20.98 & 20.81 & 20.80 &  &  &  &  &  \\*
 & B & 21.56 & 20.88 & 20.75 & 20.67 &  &  &  &  &  \\*
 & G & 23.81 & 21.76 & 21.19 & 20.61 & 1.01 & 0.60 & 0.51 & 0.45 & 0.59 \\[3pt]
\mbox{J0527$-$6132} & A & 20.33 & 20.01 & 20.00 & 19.60 &  &  &  &  &  \\*
 & B & 21.52 & 21.47 & 21.82 & 22.14 &  &  &  &  &  \\*
 & G & 23.03 & 22.31 & 23.16 & $\ldots$ & 5.50 & 5.50 & 0.44 & $\ldots$ & 0.21 \\[3pt]
\mbox{J0527$-$6136} & A & 21.60 & 21.65 & 21.89 & 22.18 &  &  &  &  &  \\*
 & B & 21.71 & 21.78 & 21.96 & 22.15 &  &  &  &  &  \\*
 & G & 22.46 & 21.22 & 20.71 & 20.31 & 1.28 & 0.82 & 0.78 & 0.72 & 0.82 \\[3pt]
\mbox{J0539$-$6051} & A & 23.07 & 22.98 & 23.03 & 23.13 &  &  &  &  &  \\*
 & B & 23.32 & 23.18 & 23.26 & 23.72 &  &  &  &  &  \\*
 & G & 21.99 & 20.92 & 20.55 & 20.24 & 0.31 & 0.25 & 0.25 & 0.25 & 0.87 \\[3pt]
\mbox{J0546$-$6132} & A & 22.81 & 22.76 & 22.71 & 22.89 &  &  &  &  &  \\*
 & B & 23.23 & 23.06 & 23.23 & 23.46 &  &  &  &  &  \\*
 & G & 22.84 & 20.71 & 20.22 & 19.75 & 0.44 & 0.44 & 0.43 & 0.38 & 0.66 \\[3pt]
\mbox{J0548$-$6116} & A & 23.28 & 23.24 & 23.33 & 23.88 &  &  &  &  &  \\*
 & B & 24.20 & 24.14 & 24.21 & 24.30 &  &  &  &  &  \\*
 & G & 21.94 & 20.72 & 20.39 & 20.03 & 0.25 & 0.25 & 0.25 & 0.25 & 0.30 \\[3pt]
\mbox{J0600$-$4356} & A & 21.62 & 21.10 & 20.97 & 21.00 &  &  &  &  &  \\*
 & B & 22.38 & 21.95 & 21.93 & 21.97 &  &  &  &  &  \\*
 & G & 21.42 & 20.40 & 19.96 & 19.66 & 1.00 & 0.88 & 0.84 & 0.78 & 0.25 \\[3pt]
\mbox{J0602$-$3852} & A & 22.39 & 22.12 & 22.18 & 22.13 &  &  &  &  &  \\*
 & B & 22.70 & 22.31 & 22.27 & 22.16 &  &  &  &  &  \\*
 & C & 21.46 & 21.06 & 21.21 & 20.95 &  &  &  &  &  \\*
 & D & 20.43 & 20.20 & 20.05 & 20.16 &  &  &  &  &  \\*
 & G & 22.57 & 21.64 & 20.81 & 20.27 & 1.98 & 0.46 & 0.42 & 0.41 & 0.70 \\[3pt]
\mbox{J1808$+$7607} & A & 22.67 & 22.43 & 22.19 & 22.14 &  &  &  &  &  \\*
 & B & 22.72 & 22.21 & 22.20 & 22.07 &  &  &  &  &  \\*
 & G & 23.97 & 22.31 & 21.89 & 21.34 & 0.68 & 0.60 & 0.51 & 0.48 & 0.18 \\[3pt]

\end{longtable}
\tablefoot{A--D identify the quasar images and G the foreground galaxy, following Fig.~\ref{fig:gallery}. G1 and G2 are the two galaxies in \mbox{J0401$-$4409}. For each galaxy we fitted the semi-major half-light radius independently in each band and a single $q_{\rm light}=b/a$ across the four bands. Blank cells mark quantities that apply only to galaxies. Dots mark the galaxy magnitudes and radii for bands with zero fitted galaxy flux (VIS in \mbox{J0323$-$4222} and $H$ in \mbox{J0527$-$6132}).}
\endgroup
\end{document}